\documentclass[twocolumn,showpacs,10pt,prb,floatfix,amssymb]{revtex4}
\usepackage{graphicx}

\begin{document}

\newcommand{\nx}{\textrm}

\title{Mobilities and Scattering Times in Decoupled Graphene Monolayers}

\author{H. Schmidt}
\author{T. L\"udtke}
\author{P. Barthold}
\author{R. J. Haug}
\affiliation{Institut f\"ur Festk\"orperphysik, Leibniz
Universit\"at Hannover, Appelstr. 2, 30167 Hannover, Germany\\
}
\date{\today}
\begin{abstract}

Folded single layer graphene forms a system of two decoupled monolayers being only a few Angstroms apart. Using magnetotransport measurements we investigate the electronic properties of the two layers conducting in parallel. We show a method to obtain the mobilities for the individual layers despite them being jointly contacted. The mobilities in the upper layer are significantly larger than in the bottom one indicating weaker substrate influence. This is confirmed by larger transport and quantum scattering times in the top layer. Analyzing the temperature dependence of the Shubnikov-de Haas oscillations effective masses and corresponding Fermi velocities are obtained yielding reduced values down to 66 percent in comparison to monolayers.

\end{abstract}

%72.80.Vp, 73.22.Pr when available

\maketitle
%\section{Introduction}
Since the discovery of stable two-dimensional crystals of carbon~\cite{Novoselov2004}, single, bi- and multi-layer graphene systems have been studied intensely~\cite{status,rise}.
Monolayer graphene exhibits outstanding electronic properties including a linear dispersion relation, zero gap, a half integer Quantum Hall Effect (QHE) and a Berry's phase of $\pi$.
Single-crystal (SC) bilayer consisting of two Bernal stacked planes also shows an unconventional QHE, yielding plateaus at integer values, but with a double step at zero filling factor and a Berry's phase of 2$\pi$~\cite{bilayer_mccann,kim_nature}.
In addition to these systems samples with two twisted monolayers have been investigated using magnetotransport measurements~\cite{me}, Raman spectroscopy~\cite{raman_misoriented} and scanning photocurrent microscopy~\cite{spcm}. The rotational stacking fault of the two layers with respect to Bernal stacking decouples them and the upper one is screened from an applied backgate voltage.
Especially when combined with a second gate on top, such a system could be a candidate to create two very closely lying electron and hole systems and could give rise to experiments recently discussed in theory~\cite{eh,eh2,eh-rt,drag}. Twisted samples can be fabricated from monolayer flakes folded during the preparation process using micromechanical exfoliation.
Beside this method epitaxial growth can be used to prepare graphene samples, also sometimes containing layers rotated with respect to Bernal stacking~\cite{berger_rotation}.\\
 
Here, we report on transport measurements on twisted monolayers, produced by micromechanical exfoliation. Analyzing the cyclotron masses, mobilities and scattering times deduced from the measurements, we find properties quite different from conventional monolayers and SC bilayer graphene.\\
Figure~1 shows a sketch of a two layer sample. After peeling of pieces from natural graphite~\cite{graphite} graphene is deposited on a silicon wafer covered with 330~nm of silicon dioxide. A suitable folded flake is located with the help of an optical microscope and e-beam lithography is used to structure the flake. A Hall-bar device is formed out of the graphene layers using plasma etching. After this, common contacts to both layers are formed by evaporating chromium and gold. The silicon backgate underneath the isolating silicon dioxide couples capacitively to the graphene. Applying backgate voltages from -~70~V to +~70~V, the density and type of majority charge carriers in the two layers can be continuously tuned from holes to electrons.
\\

\begin{figure}
\includegraphics[width=0.45\textwidth]{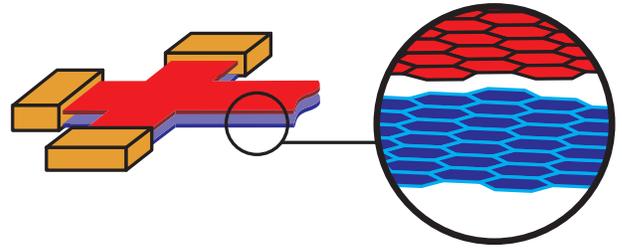}
\caption{\label{fig:fig1}Sketch of a sample containing two graphene layers, stacked and rotated with respect to Bernal stacking. A Hall-bar is etched and both layers are contacted simultaneously with chromium gold leads.}
\end{figure}

\begin{figure}
\includegraphics[width=0.45\textwidth]{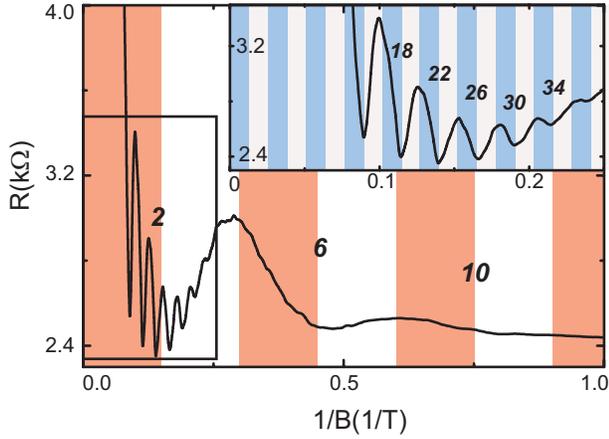}
\caption{\label{fig:fig2} Longitudinal resistance versus 1/B at a backgate voltage of $V=70$~V showing two superimposed Shubnikov-de Haas oscillations. The inset shows the oscillation with the higher carrier concentration and therefore belonging to the bottom layer. For both oscillations' minima, filling factors are marked and colored areas indicate the parts with negative slope of the corresponding oscillation.}
\end{figure}

%\section{Transport measurements}

To distinguish this system of two decoupled monolayers from a SC bilayer system, magnetotransport measurements at temperatures down to 1.5~Kelvin are performed. Applying perpendicular magnetic fields up to 13 Tesla, two sets of superimposed Shubnikov-de~Haas (SdH) oscillations are observed in the longitudinal resistance as shown in Fig.~2. Plotted over 1/B two sets of equidistant minima can be separated, both exhibiting Berry's phases of $\pi$, indicating graphene monolayers. Sweeping the backgate voltage at zero magnetic field the typical peak in the resistivity is observed indicating the charge-neutrality point. Hole doping is apparent in a shift of this peak to positive gate voltages. This could be attributed to a shift of the Fermi level due to the rotation of the layers as predicted by theory \cite{twist}. In addition the effect of impurities left over from the preparation process cannot be ruled out. Figure 3a shows this field effect at different temperatures from 1.5~K to 50~K. Two regimes can be separated: For high voltages and thus high carrier concentrations the resistivity and conductivity, respectively, don't show any temperature dependence. In contrast to this, a decrease of the resistivity is observed for increasing temperatures close to the peak. 

As shown by Morozov et al. \cite{giant} for single layer graphene, the resistivity can be separated into two parts, one due to long range scatterers, $\rho_L=1/ne\mu$, and a constant contribution due to short range scatterers. The constant part can be substracted to remove nonlinear contributions to the conductivity at high backgate voltages. 
Applying this model to the here discussed two layer system, the linearization can also be used, giving a total linearized conductivity $\sigma^*$ as plotted for 10~K in Figure 3a.
It is proportional to the carrier density at high backgate voltages and is the sum of the linearized conductivities of the two layers. 
At several fixed gate voltages SdH measurements were performed and the carrier concentrations $n_1$ and $n_2$ for the bottom and top layer were deduced. The results are shown in Fig.~3b. The carrier concentration $n_2$ is significantly lower than $n_1$ due to the screening of the electric field by the bottom layer~\cite{me}.\\

\begin{figure}
\includegraphics[width=0.45\textwidth]{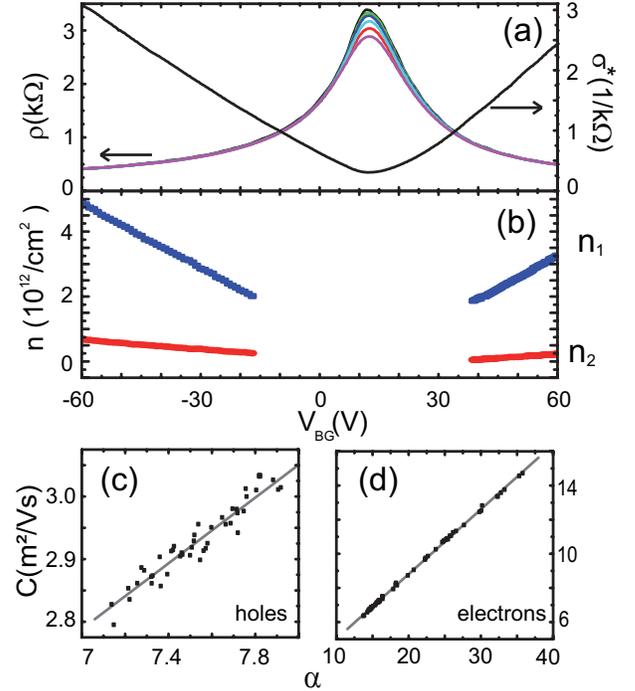}
\caption{\label{fig:fig3}a) Resistivity at zero magnetic field and different temperatures (top down: 1.5~K, 10~K, 15~K, 25~K, 35~K, 50~K). In addition, the linearized conductivity at 10~K is plotted. b) Charge carrier densities for $V<0$ (holes) and $V>0$ (electrons) and for both layers. From these, $\alpha$ and C (see text) are calculated and plotted in c) and d) with according linear fits (solid lines).}
\end{figure}

%\subsection{Mobilities}
To obtain the mobilities $\mu_1$ and $\mu_2$ of the charge carriers in the two layers, a model of two parallel conductors is used with the total conductivity being given by
\begin{eqnarray}
\label{sigma}
 \sigma(V)^*=\sigma_1(V)^*+\sigma_2(V)^*\nonumber\\
 =e\cdot(\mu_1\cdot n_1(V)+\mu_2\cdot n_2(V)).
\end{eqnarray}

Introducing $\alpha(V)$ and $C(V)$ 

\begin{eqnarray}
 C(V)=\sigma(V)^*/(e\cdot n_2(V))\nonumber\\
 \alpha(V)=n_1(V)/n_2(V) \nonumber
\end{eqnarray} 
and substituting in (\ref{sigma}) one gets a linear equation:
\begin{eqnarray}
\label{c}
   C(V)= \mu_1\cdot\alpha(V) + \mu_2.
\end{eqnarray} 

Since $\sigma(V,B=0~T)^*$, $n_1(V)$ and $n_2(V)$ are deduced from the measurements, $C(V)$ and $\alpha(V)$ can be calculated. The results are shown in Fig.~3c and 3d for the hole (V=-60 to -17V) and electron (V=39 to 60V) region, respectively. These data are in very good accordance with the expected linear behavior (\ref{c}) assuming constant mobilities. The two mobilities can be directly deduced from the slope and the offset of the according fit.
The mobility in the bottom layer yields a value of $\mu_1=$~2600~$cm^2/Vs$ for holes and 3800~$cm^2/Vs$ for electrons, respectively. Mobilities in the top layer exhibit larger values, being $\mu_2=$~9500~$cm^2/Vs$ for holes and 12300~$cm^2/Vs$ for electrons. The difference between electrons and holes in both layers is attributed to the assumed doping effects.
For both types of majority charge carriers, the mobility in the upper layer is significantly larger. This has to be explained by the fact that the second layer does not lie directly on the substrate which normally decreases the mobility of graphene due to surface charge traps, interfacial phonons and substrate stabilized ripples. 
%\subsection{Cyclotron masses}
Whereas at B=0~T for high carrier concentrations no temperature dependence of the resistivity is observed, with applied magnetic field the temperature influences the resistivity by damping of the amplitudes in the SdH oscillations. Increasing the temperature T the amplitudes of the oscillations decrease as shown in Fig.~4. This change in amplitude $\Delta R$ at a fixed magnetic field B follows the relation  $\Delta R \propto (T/B)/sinh(P*T/B)$ with the parameter $P=2\pi^2k_Bm_c/\hbar e$ containing the cyclotron mass $m_c=\sqrt{h^2n/4\pi v_F^2}$ with the Fermi velocity $v_F$.

Fitting the equation for the amplitude to our data the effective masses at different backgate voltages are obtained and shown in the inset of Fig~4. Squares indicate bottom layer values and circles the top layer ones. These values are compared to the expected masses for a monolayer with $v_F=1\cdot 10^6 m/s$, shown as gray dashed lines. The cyclotron masses in the upper layer seem to be in good accordance with this expectation although the superposition with the other oscillation and the low carrier concentration in this layer make it difficult to obtain the exact amplitude.
The values for the bottom layer show the expected linear behavior over $\sqrt{n}$ but with larger values. These increased effective masses for holes and electrons correspond to Fermi velocities of $0.66\times 10^6m/s$ and $0.81\times 10^6m/s$ (fits are shown as black dotted lines).
This reduction of the Fermi velocity in twisted samples has been theoretically predicted~\cite{twist}, especially for small rotation angles. Small reductions have been measured via Raman spectroscopy~\cite{fermiraman} on folded samples and reductions with values down to 70 percent have been observed in epitaxial graphene with a single conducting layer~\cite{fermiepitax}. The here measured strong reduction in the bottom layer confirms the assumption of a rotational stacking fault between the two layers of the folded system.\\

%\subsection{Scattering Times}
Using the obtained mobilities and cyclotron masses as extracted from our data, the transport scattering times corresponding to the long range scatterers are calculated using $\tau_{t}=m_{c}\mu/e$. Due to the larger mobilities the transport scattering times for the upper layer are larger by a factor of 2.4 for holes and 2.6 for electrons in relation to the ones for charge carriers in the bottom layer. Here, for the upper layer, cyclotron masses according to single monolayers Fermi-velocity have been used. Note that a possible underestimation of these masses would imply even higher transport scattering rates and also higher values for the quantum scattering times calculated later on.
\begin{figure}[t]
\includegraphics[width=0.45\textwidth]{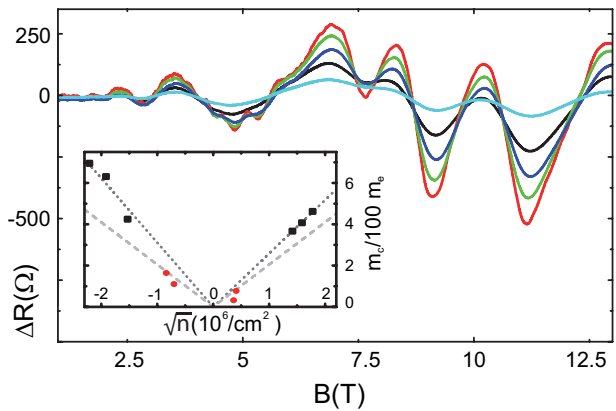}

\caption{\label{fig:fig4} Shubnikov-de Haas oscillations at -60~V backgate voltage and different temperatures (1.5~K, 10~K, 15~K, 25~K, 35~K). A background at 50~K is subtracted for better visibility. The inset shows cyclotron masses for both layers (layer 1 squares, layer 2 dots) and the values expected for a Fermi velocity of a monolayer (gray dashed lines) and the velocities fitted to our bottom layer data (black dotted line).}
\end{figure}
The dependence of the SdH oscillations on magnetic field is described by another scattering time, the quantum scattering time $\tau_q$. At 1.5~K and fixed backgate voltage the amplitudes are analyzed for different magnetic fields and additionally fitted to the Dingle factor $exp(-\pi m_c/eB\tau_q)$ to calculate these quantum scattering times \cite{tau}.
The obtained values are shown in Fig.~5 in dependence of the carrier density and compared to the transport scattering times shown as lines.
The quantum scattering times show values in the order of 25 to 70~fs for holes and electrons increasing with larger carrier density. Extrapolating the values obtained for the bottom layer down to small carrier concentrations being comparable to the ones in the top layer it becomes clear that quantum scattering times are larger in the top one. 
We assume that the upper layer is more independent from the substrate and screened from possible scattering sources. This explains why both the transport and the quantum scattering times are larger in the upper layer in respect to the bottom one.\\
%\section{Maximum resistivity}
In contrast to the high carrier concentration regime where a strong temperature dependence in the longitudinal resistance can only be observed for applied magnetic field, a monotonic and almost linear decrease of the maximum in resistivity $\rho_{max}$ by up to 15 percent is observed at B=0~T while increasing the temperature from 1.5~K to 50~K. This also persists at high magnetic fields ruling out weak localization as a possible reason. To understand this non-metallic behavior of the decoupled monolayer system it has to be compared to measurements on single monolayers with and without the influence of the substrate.
While monolayer devices on silicon dioxide show no significant temperature dependence close to the neutrality point~\cite{giant}, a decrease of the maximal resistivity has been observed in suspended graphene~\cite{npsusp}. This behavior is attributed to a very low charge carrier density near the neutrality point due to less electron and hole puddles in such samples. Theoretical calculations on the temperature dependent screening of Coulomb disorder~\cite{nptheo} predict that if the charge carrier density is small enough, a non-metallic behavior is expected, which explains the mentioned measurements on suspended graphene. We attribute the observed temperature dependence of our sample to a very low carrier concentration at the charge neutrality point for the upper layer, since this one exhibits properties comparable to suspended samples. The upper layer has reduced substrate contact, increased mobilities and scattering times, suggesting low inhomgeneity and therefore less charge carriers near the neutrality point as assumed for suspended samples\cite{suspended}.
\begin{figure}[t]
\includegraphics[width=0.45\textwidth]{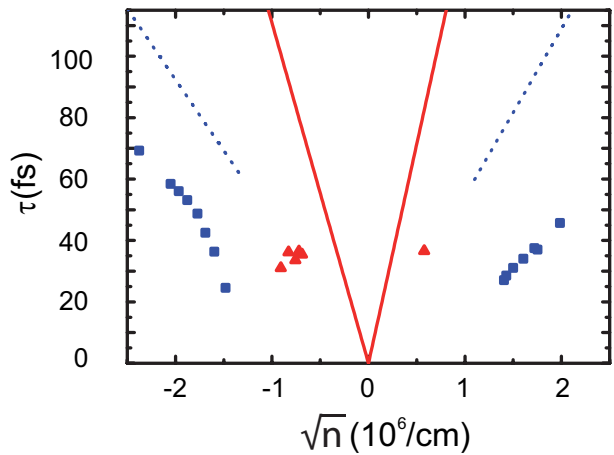}
\caption{\label{fig:fig5} Quantum scattering times in the bottom (top) layer, shown as squares (triangles) over $\sqrt{n}$. Negative values of $\sqrt{n}$ correspond to holes. The transport scattering times are calculated from the experimentally determined mobilities and masses and shown as dotted (bottom layer) and straight (upper layer) lines.}
\end{figure}

%
%\section{conclusion \& outlook} 

In summary, electronic transport in decoupled graphene monolayers has been studied in detail, showing properties quite different from single-crystal mono- and bilayer systems. Both layers act as single monolayers conducting in parallel and both show magneto oscillations with a Berry's phase of $\pi$. 
The rotational stacking fault of the two layers give rise to reduced Fermi velocities down to 66 percent of the monolayer value.
Due to the stacking, the top layer is screened from the substrate and thus has lower carrier concentrations. The majority charge carriers in this screened layer exhibit increased mobilities and scattering times. A non-metallic temperature dependence additionally indicates the unique nature of this system.\\
The authors thank V.~I.~Fal'ko and E.~McCann for discussions. This work has been financially supported by the excellence cluster QUEST within the German Excellence Initiative.

\end{document}